# First Demonstration of the Active-Mode CAOS Camera

Nabeel A. Riza (FIET) and Mohsin A. Mazhar

For the first time, demonstrated is the active-mode CAOS (i.e., Coded Access Optical Sensor) camera. The design demonstrated uses a hybrid approach to both optical device engagement and time-frequency CAOS mode operations. Specifically, time-frequency modulation of both the target illumination light source and the Digital Micromirror Device (DMD) combine to deliver the Frequency Modulation (FM)-Code Division Multiple Access (CDMA) mode of the CAOS camera. Using a 39.6 Klux white light 32 KHz FM LED source combined with a 1 KHz bit rate 4096 bits Walsh sequence CDMA code via the DMD, achieved is 58 x 70 CAOS pixels near 60 dB linear Dynamic Range (DR) imaging of a 36 patch calibrated high DR white light target. Applications for the active-mode CAOS camera are numerous and includes indoor full spectrum food inspection where a linear DR camera can play an important role for accurate measurements.

*Introduction:* Certain scenarios across various applications require a full spectrum high linear DR camera to enable robust decision making to meet critical system requirements. These applications with spectral wavelengths over 400 nm to 2500 nm and a linear DR ≥ 60 dB include industrial testing during food inspection to laser manufacturing operations to robotic vision systems for precision guidance control. Although CCD and CMOS silicon sensor-based camera systems dominate the visible band landscape [1,2], recent experiments in the visible band indicate a limited recovery of low contrast targets embedded within a high (e.g., 95 dB) linear DR scene [3]. To meet the need for full spectrum coverage from UV to visible to Near-IR with high linear DR image recovery, proposed is the CAOS camera that works on the principles of the extreme DR multi-access RF wireless mobile network. The CAOS camera design engages cost effective optical and electronic technologies that for example includes electronic Digital Signal Processing (DSP) modules, high speed Analog-to-Digital Converters (ADC), high speed point Photo-Detectors (PDs), and DMD light modulation technology [4-5]. Recent experiments have indeed shown the extreme linear DR imaging capability of the CAOS camera, reaching a 177 dB DR level [6]. Furthermore, CAOS camera experiments conducted so far operate using the passive mode of the CAOS camera where the camera images unmodulated direct or indirect target scattered light. For example, CAOS camera imaging of the Sun that is a natural unmodulated white light emitter [7].

Another powerful mode of the CAOS camera is the active light operations mode where a target to be imaged is illuminated by a CAOS time-frequency modulated coded light signal [4-5]. This *active-mode CAOS camera* forms a highly versatile linear high DR imaging platform with a precision controlled active light illumination source where depending on the imaging environment, the light power, exposure time and spectral content can be adjusted for optimal Signal-to-Noise Ratio (SNR) spectral response target imaging. This additional light control capability, whether indoor or outdoor can be highly useful in optimizing image recovery, including the use of infrared sources such as eye safe >1400 nm light. Hence, this letter presents the first demonstration of the active-mode CAOS camera. Specifically, a white light LED array target illumination source is engaged with a DMD-based CAOS camera design to demonstrate linear DR imaging of a high DR calibrated white light target. Explained next are the details of the camera experiment and its imaging results.

*Methods:* Fig.1 shows the top view of the demonstrated active-mode CAOS camera design. An external light source that is part of the proposed active-mode CAOS camera system is pointed in the direction of the target to be imaged. This active time-frequency coded light source can take a variety of forms such as LED, laser, LED array, laser array, hybrid laser-LED array sources including fiber assembled and fed light sources that can be spatially controlled and time-frequency modulated. The target scattered/reflected/transmitted light travels to the CAOS camera A1 aperture and F1 filter before entering the imaging lens L1 that maps the target plane on to the DMD plane. Depending on the imaging scenario needs such as the required linear DR, the CAOS camera can operates with various time-frequency CAOS modes [7]. For the active-mode CAOS camera, the FM-CDMA mode is engaged where the external light source provides the FM coding while the DMD provides the CDMA coding for the simultaneously viewed M CAOS pixels. It is important that the FM rate $f_C$ Hz is greater than the CDMA bit rate $f_B$ Hz so that multiple FM carrier cycles occur in a bit time to allow RF spectral processing of the FM carrier via the DSP Fast Fourier Transform (FFT) algorithm. A best practice design would be $f_C = f_B P$, where P is an integer. Because the DMD micromirrors operate as a two tilt state device, imaged light gets directed via spherical mirrors SM1 and SM2 to the point PDs labelled as PD1 and PD2. Imaging is preserved between the DMD and PD planes. The received FM-CDMA encoded RF signals from the point PDs carry the observed image M CAOS pixel irradiance data that enters a controller and processing unit containing ADC channels as well as DSP processors. Decoding of the FM-CDMA data via spectral and correlation processing recovers the viewed image [8,9]. On a design note, A1 and F1 can be electronically controlled and PD1 and PD2 can cover different spectral bands, e.g., PD1 a Silicon detector while PD2 a Germanium or InGaAs detector, giving 400 nm to 1700 nm full spectrum camera coverage [10].

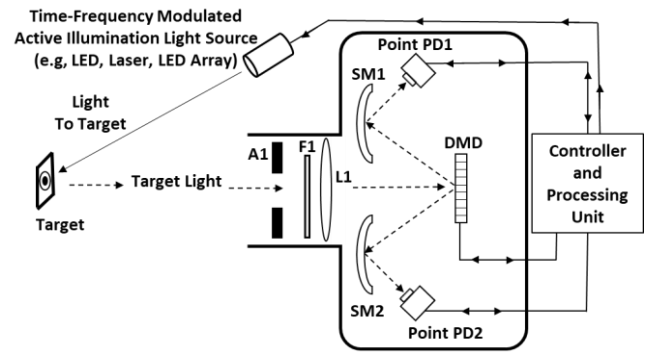

**Fig. 1** *Top view of the active-mode CAOS camera.*

*Experiments and Discussion:* The Fig.1 active-mode CAOS camera is built in the laboratory using the following components. Image Engineering (Germany) provided Custom 160 dB DR 36 patch target (see Fig.2 for design) and LG3 white light LEDs target illumination source, Vialux (Germany) DMD model V-7001, DELL 5480 Latitude laptop for control and DSP, National Instruments 16-bit ADC model 6366, and Thorlabs components that include silicon point PDs model PDA100A2, 5.08 cm diameter Iris A1, and 5.08 cm diameter 10 cm focal length front imaging lens L1, 5.08 cm diameter spherical mirrors SM1 and SM2 with 3.81 cm focal lengths. Inter-component distances are: 10.4 cm L1: DMD, 9.8 cm SM1/SM2:DMD, 6.3 cm SM1/SM2:PD1/PD2 and 261 cm target/L1. A factor of 25.1 demagnification takes place between target and DMD plane. F1 filter is not engaged for the experiment given gray-scale white light image target parameters. As shown in Fig.3, the deployed FM code within the CAOS FM-CDMA mode is a $f_C$ = 32 KHz and 50% duty cycle square wave signal driving the white light LEDs in LG3. The LG3 active source provides 39.5 Klux that illuminates the gray-scale calibrated 36 patch target. The CDMA mode via the DMD uses an $f_B$ = 1 KHz giving P=32 cycles per CDMA code bit with a 4096 bits Walsh sequence providing M= 58 x 70 CAOS pixels grid in the viewed image space.

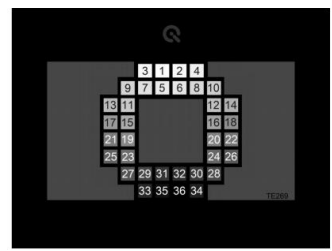

**Fig. 2** *Custom 36-Patch high DR target design with numbered patches.*

The ADC sampling rate is 1 MSps with the point PD electronic gain set to 40 dB. The digitized FM-CDMA signal from the CAOS camera point PD1 first undergoes a N=1024 point FFT with a DSP FFT gain of 10log (N/2) dB and then CDMA decoding via correlation processing to recover the scaled irradiance values for the observed M = 4060 CAOS pixels. Each CAOS pixel is made of 13 x 13 micromirrors where each micromirror is 13.68 μm x 13.68 μm. A single PD channel is used for image processing although 2 channels can also be used for dual spectrum [10] or lower noise differential processing [11]. Note that DR in dB = 10 log ($P_{max}/P_{min}$), where $P_{max}$ and $P_{min}$ are the scaled maximum and minimum optical power levels measured per imaged CAOS pixel, respectively. SNR = (Scaled signal optical power of CAOS pixel)/ (Scaled noise optical power of CAOS image Dark Pixel) and an SNR = 1 produces the maximum possible camera detection DR. The scaled noise optical power signal is computed from the CAOS camera imaged pixels without illumination, i.e., the black pixels. The LG3 illumination is controlled to a level of 1/4000 or 0.25% and has a ≥ 95% uniformity in its white light illumination area. The CAOS camera captured both the white illumination zone inside the field-of-view of the camera as well as the surrounding light-free black area (see Fig.4). Data analysis of the Fig.4 image shows a measured uniformity of 95% with a measured average scaled illumination value of 0.76 on a zero to 1 scale.

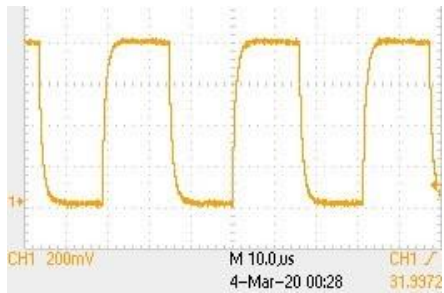

**Fig. 3** *Active CAOS camera deployed 32 KHz FM signal CAOS code for the white LEDs-based 39.6 Klux source illuminating the target.*

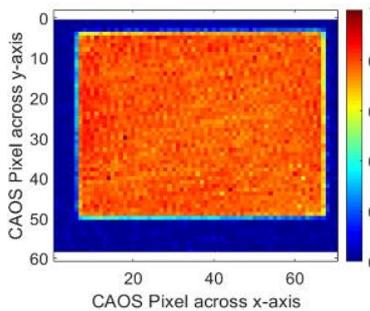

**Fig. 4** *Active CAOS camera generated image of the LG3 active light source illumination at target screen location without the 36-patch target. Image shown in a 0 to 1 linear scale.*

With the 36-patch target illuminated by the active source with a 32 KHz FM-code and 1 KHz CDMA-code, Fig.5 shows the CAOS FM-CDMA mode recovered 14 target patches indicating a 59.4 dB DR recovery with an SNR>1 (see Table 1). Fig.6 shows the image recovery is linear over this range with a fitted slope of 0.94 versus the ideal slope of 1.

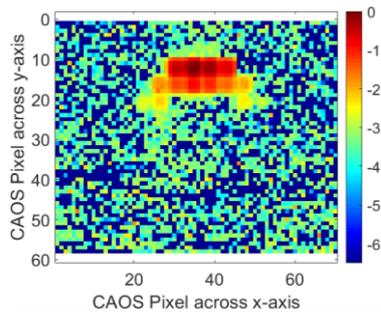

**Fig. 5** *Active CAOS Camera generated near 60 dB DR image of the Fig.2 36-patch target. The camera used its FM-CDMA mode. Image shown in log scale in order to display a high DR on a computer screen.*

| Patch # | DR (dB) | M (dB) | SNR | Patch # | DR (dB) | M (dB) | SNR |
|---|---|---|---|---|---|---|---|
| 1 | - (ref) | - | 836.7 | 8 | 32 | 30.0 | 26.6 |
| 2 | 4.6 | 4.2 | 516.8 | 9 | 36.6 | 34.0 | 16.8 |
| 3 | 9.2 | 8.9 | 301.9 | 10 | 41.2 | 37.5 | 11.2 |
| 4 | 13.8 | 12.5 | 198.3 | 11 | 45.8 | 44.5 | 5.0 |
| 5 | 18.2 | 17.1 | 117.4 | 12 | 50.2 | 48.4 | 3.2 |
| 6 | 22.8 | 21.5 | 70.7 | 13 | 54.8 | 52.4 | 2.0 |
| 7 | 27.4 | 26.0 | 42.1 | 14 | 59.4 | 55.3 | 1.4 |

**Table 1:** *Active-mode CAOS camera measured target patch DR values.*

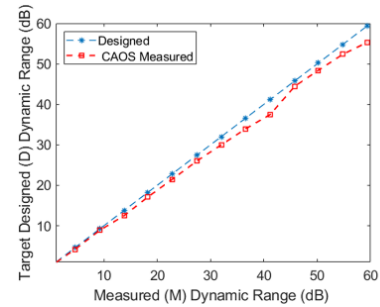

**Fig. 6** *Linearity plot for the active-mode CAOS camera image data.*

*Conclusion:* For the first time, demonstrated is the active-mode CAOS camera that is empowered by an external light source that implements part of the CAOS camera light encoding operations. The active-mode CAOS camera features additional flexibility to optimize linear DR imaging via illumination spatial and temporal/frequency control. Experiments show near 60 dB linear DR white light imaging via LED-based illumination that implements the FM-mode while the DMD that implements the CDMA-mode of the CAOS camera to complete CAOS FM-CDMA mode imaging of a high DR calibrated target. The active-mode CAOS camera can be used for various indoor and out-door imaging applications that require a linear DR and full spectrum coverage within one cost-effective camera unit.

N. A. Riza and M. A Mazhar (*School of Engineering, University College Cork, College Road, Cork, Ireland*)

E-mail: n.riza@ucc.ie